# ON THE DYNAMICS AND THERMODYNAMICS OF SMALL MARKOW-TYPE MATERIAL SYSTEMS

Andrzej Trzęsowski

Department of Theory of Continuous Media, Institute of Fundamental Technological Research, Polish Academy of Sciences, Pawińskiego 5B, 02-106 Warsaw, Poland
e-mail adresses: atrzes@ippt.gov.pl , artrzes@gmail.com

**Abstract.**
The collective properties of small material systems considered as semidynamical systems revealing the Markov-type irreversible evolution, are investigated. It is shown that these material systems admit their treatment as thermodynamic systems in diathermal and isothermal conditions. A kinetic equation describing statistical regularities of the Markov-type material systems and constrained by the compatibility condition with the first and second laws of thermodynamics and with the relaxation postulate, is proposed. The influence of external parameters on the stationary states of small material systems endowed with their own energy independent of dynamics is discussed.

## 1. Introduction

The problem of comprehensive description of collective properties of material systems, observed not on the macroscopic scale but on different mesoscales, appears in the theory of nanostructured materials [1, 2, 3]. Particularly, it is indispensable to clarify the applicability of the concepts of thermodynamics to material systems on the small length scales [2]. For example, if we are dealing with the macroscopic observation level scale, then the standard procedure to show the existence of thermodynamic limit, and therefore temperature, is based on the idea that, as the spatial extension increases, the surface of a region in space grows slower than its volume [2]. However, it is not the case of small material systems [1, 3]. The aim of this paper is to show that the method of description of collective properties of material systems, which has been formulated in [4] for stochastic systems with the countable space of states, can be helpful, after its modification and generalization (based on the theory of semidynamical systems and on the statistical theory of stationary states), for better understanding of the dynamics and nonequilibrium thermodynamics of small material systems with the state space being countable or having the cardinality of continuum. The considered material systems can be deterministic (with the dynamics described by semidynamical systems) or stochastic but admit the Markov-type evolution and admit their treatment as thermodynamic systems in diathermal and isothermal conditions (Sections 3 - 5). In Section 2 are reviewed briefly different methods of formulating a probabilistic representation of dynamical (or semidynamical) systems, a kinetic equation, named the Kolmogorov-type kinetic equation, is introduced, and the Markov-type evolution processes of nonsingular semidynamical systems are defined. In Section 3 are considered not closed Markov-type small material systems and the condition that the Gibbs distribution fulfills the Kolmogorov-type kinetic equation is given. In Section 4 the influence of external parameters on the stationary states of small material systems endowed with their own energy independent of dynamics is discussed. Moreover, in Section 4, the notion of conditional entropy is considered and the relationship between the existence of positive absolute temperature and the dependence of entropy on the internal energy of the system is presented. In Section 5 are formulated conditions of the thermodynamical admissibility of Markov-type evolution processes of the considered material systems. The proposed method of the description of collective properties of these systems admits its consistency with the so-called Prigogine's selection rule for irreversible dynamical processes (Section 6). In the paper the *reversibility* of deterministic or stochastic dynamical processes (see Section 2) means that when the direction of time is reversed the behavior of these processes remains the same.

## 2. Probabilistic representations of dynamical systems

Let us consider a material system that dynamical behavior is defined by a topological space $X$ (countable or having the cardinality of continuum) of all its admissible states and by a *deterministic topological dynamical system* [5] defined as a continuous Abelian semigroup or group $G = \{S_t : X \to X, t \in T\}$ of continuous transformations, where $T = R = (-\infty, +\infty)$ or $T = R_+ = [0, +\infty)$ (with the internal operation $t_1 \pm t_2 \in T$ for every $t_1 \geq t_2 \in T$) for the group or semigroup, respectively, and

$$S_t \circ S_s = S_{t+s}, \quad S_0 = \mathrm{id}_X, \quad s, t \in T. \tag{2.1}$$

We can also consider a semigroup G with $T = R_- = (-\infty, 0]$ as the semigroup of parameters. If $x_0 \in X$ is a distinguished state of the material system, called its *initial state*, then

$$x_t = \mathrm{x}(x_0; t) \equiv S_t x_0 \tag{2.2}$$

denotes the *instantaneous state* of this system at the instant $t \in T$. For example, if $X$ is a differential manifold and $\mathrm{x}(x_0; \cdot): T \to X$ is the general solution of the differential equation

$$\dot{\mathrm{x}} = \mathbf{v}(x), \quad \mathrm{x}(0) = x_0, \tag{2.3}$$

where $\dot{\mathrm{x}} = d\mathrm{x}/dt$ and $\mathbf{v}$ is a vector field on $X$ tangent to $X$, then Eq.(2.2) can be considered as a definition of a (deterministic) smooth dynamical system generated by the differential equation of Eq.(2.3) [5, 6].

We assume that the space $X$ is additionally endowed with a distinguished σ-finite, nonnegative measure $\mu : \Gamma \to R_+$ where $\Gamma$ denotes a σ-algebra of subsets of $X$ [6]. In particular, if $X$ is a countable set ($\mathrm{card}\, X \leq \aleph_0$) and as $\Gamma$ is taken the set of all subsets of $X$, then $\mu$ is the so-called *counting measure* on $X$ defined for $A \in \Gamma$ by the rule:

$$\mu(A) = \begin{cases} \text{card } A & \text{for} \quad \text{card } A < \aleph_0, \\ \infty & \text{for} \quad \text{card } A = \aleph_0. \end{cases} \qquad (2.4)$$

Further on a measure space $X_\mu = (X, \Gamma, \mu)$ [6] is called the *state space* of a system. The space $X_\mu$ is also called a *phase space* of the system. By $L_\mu(X)$ we will denote the linear Banach space of $\mu$-measurable functions $f: X \to \mathrm{R}$ such that [6]:

$$\|f\| = \int_X |f(x)| \mathrm{d}\mu(x) < \infty. \qquad (2.5)$$

Particularly, if $\mu$ is the counting measure, then

$$\|f\| = \sum_{x \in X} |f(x)| < \infty. \qquad (2.6)$$

Note that $L_\mu(X)$ is the so-called Banach algebra with respect to the internal pointwise multiplication of functions belonging to this space [6]. Now, the semigroup (or group) G of transformations $X_\mu$ should consist of measurable transformations [5], that is the so-called condition of *double measurement* of these transformations should be fulfilled [7]:

$$\forall A \in \Gamma, \forall t \in T, \left(\mathrm{S}_t(A), \mathrm{S}_t^{-1}(A) \in \Gamma\right). \qquad (2.7)$$

In the literature a deterministic (and topological or smooth) dynamical system defined as the semigroup $\mathrm{G} = \{\mathrm{S}_t : X_\mu \to X_\mu, t \in T\}$ fulfilling the condition of double measurement is frequently called a *semidynamical system* [7].

The most prominent example of dynamical systems generated by differential equations and fulfilling the above condition of double measurement is the *Hamiltonian dynamical system* describing the dynamics of a material system consisting of $N$ identical particles. In this case $X = \mathrm{R}^{2n}$, $n = 3N$, $\mu$ is the Lebesgue measure, and Eq.(2.3) with

$$\mathbf{v}(x) = \mathbf{J} \nabla_x H(x),$$
$$\mathbf{J} = \begin{bmatrix} \mathbf{0} & \mathbf{I} \\ -\mathbf{I} & \mathbf{0} \end{bmatrix}, \quad \mathbf{I} = \mathrm{diag}(1, 1, ..., 1) \in \mathrm{GL}(n, \mathrm{R}), \qquad (2.8)$$

where $\mathrm{GL}(n, \mathrm{R})$ is the group of nonsingular $n \times n$ real matrices, $H: X \to \mathrm{R}$ is the Hamiltonian of the system and $\nabla_x$ denotes the gradient operator with respect to variables $x \in X$, is considered [8]. Let G denote the Hamiltonian dynamical system defined by Eqs.(2.1)-(2.3) and (2.8). It can be shown that G is the one-parameter *group* of transformations of the space $X = \mathrm{R}^{2n}$, called *canonical transformations*, that preserve the Lebesgue measure $\mu$ (that is preserve the volume in $X$) [9, 10].

If the initial state $x_0$ of a material system is known up to the probability of its localization in a subset $A \in \Gamma$:

$$P(x_0 \in A) \equiv P(A) = \int_A p(x)\,d\mu(x),$$
$$p \in L_\mu(X), \quad p \geq 0, \quad P(X) = 1, \tag{2.9}$$

then the probability that an instantaneous state of the system is localized at the instant $t \in T$ in the set $A$ can be defined as:

$$P(x_t \in A) \equiv P_t(A) = \int_A p_t(x)\,d\mu(x) = P\left(x_0 \in S_t^{-1}(A)\right). \tag{2.10}$$

Particularly, it results from the definition of canonical transformations, that

$$p_t = p \circ S_t^{-1} \in D_\mu(X), \tag{2.11}$$

where it was denoted:

$$D_\mu(X) = \left\{ p \in L_\mu(X): \; p \geq 0, \; \int_X p(x)\,d\mu(x) = 1 \right\}. \tag{2.12}$$

It follows that the Hamiltonian dynamical system (2.8) generates, according to Eqs.(2.2), (2.3), (2.9)-(2.12), a *semigroup* $U = \left\{ U_t : D_\mu(X) \to D_\mu(X), \; t \in T \right\}$ of transformations acting according to the rule:

$$U_t p = p \circ S_t^{-1},$$
$$p = p_0, \quad U_0 = \mathrm{id}_{D_\mu(X)}, \tag{2.13}$$

extensible to the linear mappings $U_t : L_\mu(X) \to L_\mu(X)$, and such that the sufficiently smooth probabilistic densities of Eq.(2.11) fulfill the so-called *Liouville equation*:

$$\partial_t \bar{p}(x, t) = L\bar{p}(x, t),$$
$$\bar{p}(x, t) = p_t(x), \quad \bar{p}(x, 0) = p(x), \tag{2.14}$$

where $L$ is the Liouville operator acting according to the rule

$$Lf = \{f, H\} = -\{H, f\},$$
$$f, H \in C^k(X), \quad k \geq 1, \tag{2.15}$$

and $\{\cdot,\cdot\}$ denotes the Poisson brackets [9]. Note that since in order to solve the Liouville equation we ought to known, in general, a solution of the Hamiltonian equations, it is the problem unrealizable for macroscopic systems. Consequently, in the statistical physics are considered approximate solutions of the Liouville equation that describe statistical regularities of the Hamiltonian system [9].

If we are dealing with the *oriented in time* evolution of a *stochastic material system* with the state space $X_\mu$, then the randomness of instantaneous states of the system can be described by a family $x_T = \{x_t : t \in T\}$ of mappings

$$\begin{aligned} x_t : (\Omega, P) &\to X_\mu, \quad t \in T, \\ P : \Omega &\to [0,1], \quad P(\Omega) = 1, \end{aligned} \quad (2.16)$$

where $(\Omega, P)$ denotes a probabilistic space of elementary events and $T\ (= R_+ \text{ or } R_-)$ is an one-parameter, additive and Abelian semigroup. If

$$P(x_t \in A) \equiv P(\omega \in \Omega : x_t(\omega) \in A) = \int_A p_t(x)\,d\mu(x), \quad (2.17)$$

then we can assume the existence of an Abelian semigroup $U = \{U_t, t \in T\}$ of mappings

$$U_t : D_\mu(X) \to D_\mu(X), \quad U_0 = \mathrm{id}_{L_\mu(X)}, \quad (2.18)$$

where Eq.(2.12) is taken into account, such that if $x_0$ is a random variable with a probability density function $p$ of Eq.(2.9), then

$$\forall t \in T,\ U_t p = p_t, \quad (2.19)$$

and these mappings are extensible to the linear operators $U_t : L_\mu(X) \to L_\mu(X)$. It ought to be stressed that the way in which the probabilistic representation of a dynamical system is introduced depends on the kind of randomness associated with the dynamics. For example, in the case of a deterministic dynamical system of Eq.(2.3), the only way to introduce its probabilistic representation is the randomness of its initial conditions. Another situation takes place e.g. in quantum systems, in which the randomness can be related with the Heisenberg uncertainty relation.

More generally, we can consider a *semigroup* U of linear operators in $L_\mu(X)$ fulfilling the condition (2.18) but not necessarily generated by the family $x_T$ of random variables. If these linear operators are the so-called *Markov operators*, that is:

$$\forall f \in L_\mu(X), \left(f \geq 0 \Rightarrow \forall t \in T, \left(U_t f \geq 0 \text{ and } \|U_t f\| = \|f\|\right)\right), \quad (2.20)$$

then U is called a *stochastic semigroup* [7]. It follows from the condition (2.20) that should be [7]:

$$\forall f \in L_\mu(X), \forall t \in T, \|U_t f\| \leq \|f\|. \quad (2.21)$$

Thus, the Markov operators are contractions. Let $\pi \in D_\mu(X)$ be the so-called *stationary density*, defined by [7]:

$$\forall t \in T, \ U_t \pi = \pi. \tag{2.22}$$

A stochastic semigroup is called *asymptotically stable* if there exists exactly one stationary density $\pi$ such that [7]

$$\forall p \in D_\mu(X), \ \lim_{t \to \infty} \|U_t p - \pi\| = 0. \tag{2.23}$$

For example, let us consider a family $K = \{K_t : X_\mu \times X_\mu \to \mathbb{R}, \ t \in T\}$ of the so-called *stochastic kernels* satisfying, for every $s, t \in T$ and almost everywhere on $X_\mu$, the so-called *Chapman-Kolmogorov* equation [7, 11]:

$$\begin{aligned} K_{t+s}(x, y) &= \int_X K_t(x, z) K_s(z, y) \mathrm{d}\mu(z), \\ K_t(x, y) &\geq 0, \quad \int_X K_t(x, y) \mathrm{d}\mu(y) = 1. \end{aligned} \tag{2.24}$$

Given $K$ we can define a stochastic semigroup by setting for any $f \in L_\mu(X)$:

$$U_t f(x) = \int_X K_t(x, y) f(y) \mathrm{d}\mu(y). \tag{2.25}$$

Note that for every $t_0, t \in T$, $t > t_0$, we have [7]:

$$\begin{aligned} \forall p \in D_\mu(X), \ U_t p(x) &\geq h_0(x), \\ h_0(x) &= \inf_{z \in X} K_{t_0}(x, z), \quad x \in X. \end{aligned} \tag{2.26}$$

It can be shown [7] that if $K$ is a family of stochastic kernels such that

$$\int_X h_0(x) \mathrm{d}\mu(x) > 0 \tag{2.27}$$

for some $t_0 \in T$, then the semigroup defined by Eq.(2.25) is asymptotically stable. The probabilistic interpretation of the above stochastic semigroup is defined by Eqs.(2.16)-(2.19) and (2.25). For example, it is the case of *Markov chains* (i.e. Markov processes with the countable state spaces [11]) with the continuous time which has been considered in [4]. Farther on a stochastic semigroup generated by a family of stochastic kernels is called the *Chapman-Kolmogorov semigroup*.

Stochastic semigroups corresponding to the randomness of instantaneous states of material systems (i.e., defined by Eqs.(2.16)-(2.20)) appear mainly in pure probabilistic problems such as random walks, stochastic differential equations and many others (e.g. in the problem of Markovian description of collective properties of sys-tems with the countable state space [4]). However, they can all be generated also by deterministic semidynamical systems [7]. A semidynamical system is called *nonsingular* if in addition

$$\forall A \in \Gamma, \left[ \mu(A) = 0 \Rightarrow \forall t \in T, \mu(S_t(A)) = \mu(S_t^{-1}(A)) = 0 \right]. \tag{2.28}$$

For any nonsingular semidynamical system G we can univocally define the *stochastic semigroup* $U \equiv U[G] = \{U_t : L_\mu(X) \to L_\mu(X), t \in T\}$ assuming that for every measurable set $A \subset X$, we have (cf. Eqs.(2.9) and (2.10)):

$$\forall f \in L_\mu(X), \forall t \in T, \int_A U_t f(x) d\mu(x) = \int_{S_t^{-1}(A)} f(x) d\mu(x). \tag{2.29}$$

The such defined stochastic semigroup fulfills additionally the following condition:

$$\mathrm{supp}(U_t f) \subset S_t(\mathrm{supp}\, f), \tag{2.30}$$

where $\mathrm{supp}\, f = \{x \in X : f(x) \neq 0\}$ is the support of $f$ [7]. It follows from Eq.(2.30) that if $A = \mathrm{supp}\, f$, then $U_t f(x) = 0$ for $x \notin S_t(A)$ [7]. A nonsingular semidynamical system G is called *statistically stable* if the corresponding stochastic semigroup is asymptotically stable. The behavior of U[G] allows to determine many properties of the semidynamical system G. For example, let us consider the problem of the existence of a measure $\mu_0$ *invariant* under G, that is such that [7]

$$\mu_0(S_t^{-1}(A)) = \mu_0(A) \tag{2.31}$$

for every measurable set $A \subset X$ and $t \in T$ (as e.g. in the case of Hamiltonian dynamical systems). Assume now that a measure $\mu_0$ is normalized ($\mu_0(X) = 1$) and invariant under G. The pair $(X_{\mu_0}, G)$ is called *exact* if for every measurable set $A \subset X$ the following condition is fulfilled:

$$\mu_0(A) > 0 \Rightarrow \lim_{t \to \infty} \mu_0(S_t(A)) = 1. \tag{2.32}$$

Let G be a nonsingular semidynamical system and let U[G] denotes the stochastic semigroup associated with its. If $f \in L_\mu(X)$, then the measure

$$\mu_f(A) = \int_A f(x) d\mu(x), \quad A \in \Gamma, \tag{2.33}$$

is invariant under G if and only if

$$\forall t \in T, \, U_t f = f, \tag{2.34}$$

where Eq.(2.29) was taken into account [7]. Moreover, if the semigroup U[G] is asymptotically stable, $f \in D_\mu(X)$ is its unique stationary density and $\mu_f$ is the measure given by Eq.(2.33), then the pair $\left(X_{\mu_f}, G\right)$ is exact and $\mu_f$ is the unique absolutely continuous normalized (nonnegative) measure invariant under G [7].

The probabilistic interpretation of a nonsingular semidynamical system G can be formulated if the corresponding stochastic semigroup U[G] is consistent with a stochastic process defined by Eqs.(2.16)-(2.19). It can be e.g. the case of *Markov processes* [11]. We will assume additionally, generalizing the case of Markov chains [4, 11], the existence and finiteness of the so-called *transition intensities* $W(x,y) \geq 0$, $x \neq y$, from the state $x$ to the state $y$ and the so-called *exit intensities* from the states $x$ of the system:

$$w(x) = \int_X W(x,y) \, d\mu(y) > 0. \tag{2.35}$$

Then, the probability density $p_t$ defined by Eqs.(2.19) can be assumed in the form of $p_t(x) = \bar{p}(x,t)$ where $\bar{p}: X \times T \to \mathbb{R}_+$ is a solution of the following version of the so-called *Kolmogorov equation* considered in the theory of Markov processes:

$$\partial_t \bar{p}(x,t) = -w(x)\bar{p}(x,t) + \int_X W(y,x)\bar{p}(y,t) \, d\mu(y),$$
$$\bar{p}(x,0) = p(x). \tag{2.36}$$

This equation has the physical meaning of a kinetic equation that defines the probabilistic representation of the Markov-type evolution of a material system on the basis of the balance of the intensities of reaching and leaving the states of this system. Therefore it can be named the *Kolmogorov-type kinetic equation*. Note that the quantity

$$\tau(x) = \frac{1}{w(x)} < \infty \tag{2.37}$$

can be interpreted as the *mean residence time* of the Markov-type evolution process in the state $x \in X$ [4].

Let us denote by $L_G$ the linear operator (in general unbounded) in the Banach algebra $L_\mu(X)$ defined as

$$L_G f(x) = -w(x)f(x) + \int_X W(y,x)f(y) \, d\mu(y). \tag{2.38}$$

Frequently, we have to take into account *constraints* restricting the state space [1] as well as concerning the evolution of the system. For example, it is easy to observe that if the exit intensities $w$ are commonly bounded

$$\exists w_0 > 0, \ \forall x \in X, \ w(x) \leq w_0, \tag{2.39}$$

then the operator $L_G$ is bounded:

$$\|L_G\| = \sup_{\|f\|\leq 1} \|L_G f\| \leq 2w_0, \qquad (2.40)$$

where Eq.(2.5) was taken into account. It follows from Eq.(2.36) that then the stochastic semigroup U[G] consists of operators of the exponential form [10]:

$$U_t = \exp(tL_G) \equiv \sum_{\acute{n}=0}^{n=\infty} \frac{t^n}{n!} L_G^n, \qquad L_G^0 \equiv I, \qquad (2.41)$$

where $If = f$ for any $f \in L_\mu(X)$.

The existence of a kinetic equation of the form (2.36) is an important fact from the point of view of the physical applications of stochastic semigroups to the description of irreversible processes. It follows from the following form of transition probabilities of the Markov processes governed by such equation [4, 11]:

$$P(x_{t+h} = y | x_t = x) = P(x_h = y | x_0 = x) = W(x,y)h + o(h),$$
$$P(x_{t+h} \neq y | x_t = x) = P(x_h \neq y | x_0 = x) = w(x)h + o(h), \qquad (2.42)$$

where $o(h)/h \to 0$ for $h \to 0$, uniformly with respect to $x \in X$ for given $y \in X$, $x \neq y$. Hence in this case the description of the *irreversible evolution* of the system can be reduced to the investigation of its behavior for short time periods, that is to the formulation of the physical hypothesis about the form of the transition probabilities. For example, it follows from Eqs.(2.39) and (2.42) that, independently of the choice of $x \in X$, should be

$$\forall y \in X, \ (y \neq x \Rightarrow P(x_h \neq y | x_0 = x) \leq w_0 h + o(h)). \qquad (2.43)$$

Further on the considered material systems (deterministic with the dynamics described by nonsingular semidynamical systems or stochastic [4, 11]) are constrained by the condition that the associated stochastic semigroups are generated by the Kolmogorov-type kinetic equation (2.36). Consequently, one can say that such material systems and such stochastic semigroups are the *Markov-type*. Note that such material systems can admit the *Markov-type irreversible evolution*.

## 3. Stationary states

Let us consider a material system with the state space $X_\mu$ and such that every state $x \in X_\mu$ has its own energy $E_x \in R_+$ independent of the dynamics of the system (and called *internal energy* of the system in the state *x*). The dynamics of the material system is described by a nonsingular semidynamical system acting in the state space $X_\mu$ (see remarks previous to Eq.(2.28)). We can introduce now the *mean internal energy* functional $E : D_\mu(X) \to R_+$ acting according to the following rule:

$$\forall p \in D_\mu(X), \ E(p) = \int_X e(x) p(x) \mathrm{d}\mu(x), \qquad (3.1)$$
$$e: X_\mu \to \mathbb{R}_+, \qquad e(x) = E_x \ \text{for} \ x \in X_\mu,$$

where Eq.(2.12) was taken into account, and we can distinguish the class of probabilistic measures giving the some value $\varepsilon$ of the mean internal energy:

$$D_{\mu,\varepsilon}(X) = \{p \in D_\mu(X) : E(p) = \varepsilon\}. \qquad (3.2)$$

Since the state space $X_\mu$ can be countable or can has the cardinality of continuum, the *distribution of energy e* is admitted to be a discrete or a continuous function of the variable *x*, respectively.

Though one can adapt the presented description of *collective properties* of material systems to the description of macroscopic systems, however first of all will interest us *small systems* (see Section 1 and the remarks following Eq.(2.15)). For example, let the system consists of the finite number $N$ of identical particles, the state space $X$ is countable and

$$E(p) = \sum_{x \in X} E_x p_x < \infty,$$
$$\sum_{x \in X} p_x = 1, \ p_x = p(x) \geq 0, \ E_x \geq 0, \qquad (3.3)$$

where Eqs.(2.4)-(2.6) and (3.1) were taken into account. Let us denote by $n_x$ the (finite) number of particles being in the state $x \in X$ endowed with the internal energy $E_x$. In the classical statistical physics, the probability $p_x$ of Eq.(3.3) that a particle of the system has the internal energy $E_x$ in the state *x*, is assumed, under certain physical conditions [9, 12], in the following form:

$$p_x = \frac{n_x}{N}, \quad N = \sum_{x \in X} n_x, \qquad (3.4)$$

and it is approximated, in the so-called *thermodynamic limit* [9, 12], by

$$p_x = \lim_{N \to \infty} \left(\frac{n_x}{N}\right). \qquad (3.5)$$

The approximation is the better, the greater $N$ is. However, the condition (3.5) can not be accepted in the case of small material systems.

In statistical physics is considered the so-called *Boltzmann entropy* functional $S: D_\mu(X) \to \mathbb{R}$ defined by:

$$\forall p \in D_\mu(X), \ S(p) = \int_X \mathrm{s}(p(x)) \mathrm{d}\mu(x), \qquad (3.6)$$

where

$$s(z) = \begin{cases} -k_B z \ln z & \text{for } z > 0, \\ 0 & \text{for } z = 0, \end{cases} \quad (3.7)$$

and $k_B$ is the Boltzmann constant. This functional is treated, in the thermodynamic limit, as a measure of the statistical information concerning the energetic states of macroscopic systems [9, 12]. Nevertheless, it can be accepted also as a measure of uncertainty in the statistical description of processes in *microscopic bodies* [9]. This measure of information takes its maximum value consistent with the fixed value of the mean energy,

$$\exists \pi \in D_{\mu,\varepsilon}(X), \ \forall p \in D_{\mu,\varepsilon}(X), \ S(p) \leq S(\pi), \quad (3.8)$$

on the so called *Gibbs distribution* $\pi$ of the following form:

$$\pi(x) = Z^{-1} \exp\left(-\frac{\beta E_x}{k_B}\right),$$
$$Z = \int_X \exp\left(-\frac{\beta E_x}{k_B}\right) d\mu(x) < \infty, \quad (3.9)$$

where $\beta > 0$ is a constant. For sufficiently smooth distributions, equality in Eq.(3.8) implies that $p = \pi$ almost everywhere on *X*. It should be stressed that, in the framework of classical statistical physics, the Gibbs distribution is applied only in the thermodynamic limit, that is, it should be understood then in the sense of Eq.(3.5) with $p_x = \pi(x)$. The probabilistic representation $(X_\mu, \pi)$ defined in this way is called *canonical ensemble* (for the countable state space) [9, 12].

Denoting

$$F(\pi) = -E_B \ln Z, \quad E_B = \theta k_B, \quad \theta = \beta^{-1}, \quad (3.10)$$

we obtain the following relation:

$$F(\pi) = E(\pi) - \theta S(\pi), \quad (3.11)$$

where $E(\pi) \equiv \varepsilon$ is a fixed mean internal energy of the system, $S(\pi)$ is the maximal entropy corresponding to *E* and defined by Eqs.(3.6), (3.7) and (3.9), and the scalar $\theta > 0$ defines, according to Eq.(3.10), the characteristic energy $E_B$ of the system. If it is a *macroscopic* and *closed* system (i.e. the material system is energetically isolated), then the scalar $\theta$ can be identified with the *absolute thermodynamic temperature* of the system and the quantity $F(\pi)$ can be recognized as the *free energy* of the system [9]. Moreover, for a closed system, the state of statistical equilibrium (i.e. the condition that the distribution $\pi$ is independent of time) covers with the state o*f thermodynamic equilibrium* [9].

If the system is *not closed*, then his states can be dependent on the ambient temperature [9]. Particularly, it can be the case of a system (macroscopic or microscopic) with a thermally conducting boundary (called a *diathermal boundary*), admitting thermally activated processes and coupled with its environment (a thermostat). In this case we can assume that $\theta$ covers with the temperature of the environment [4] or, more generally, that this temperature is produced by a thermostat. $F(\pi)$ takes then the physical meaning of a *generalized free energy* corresponding to the *Gibbs distribution* and this distribution can describe, in general, a *stationary nonequilibrium state* of a small material system (see remarks following Eqs.(3.7) and (3.11)). We will call this state a *Gibbs state*.

If we are dealing with a *Markov-type material system*, say with a nonsingular semidynamical system (see remarks at the very end of Section 2), then it follows from Eqs.(2.35), (3.9), (3.10) and the condition

$$\partial_t \bar{p} = 0 \tag{3.12}$$

that the Gibbs distribution fulfills the Kolmogorov-type kinetic equation (2.36) if the following analogue of the so-called condition of *microscopic reversibility* (called also the condition of *detailed balance* [12]) is satisfied:

$$\forall x, y \in X, \left( x \neq y \Rightarrow W(x,y) > 0, \ \pi(x)W(x,y) = \pi(y)W(y,x) \right). \tag{3.13}$$

This condition is fulfilled if the transition intensities $W(x,y)$ are of the form:

$$W(x,y) = q(x,y)\exp\left(\frac{E_x}{E_B}\right),$$
$$q(x,y) = q(y,x) > 0 \text{ for } x \neq y. \tag{3.14}$$

The formula (3.14) can be written, without losing generality, in the form:

$$W(x,y) = \nu \exp\left(-\frac{U_{xy}}{E_B}\right), \tag{3.15}$$

where $\nu > 0$ is a constant with the dimension of frequency, and it was denoted

$$U_{xy} = E_{xy} - E_x, \quad E_{xy} = E_{yx} \geq 0. \tag{3.16}$$

The ratio

$$k_{xy} = \frac{W(x,y)}{W(y,x)} = \frac{\pi(y)}{\pi(x)} = \exp\left(\frac{E_x - E_y}{E_B}\right) \tag{3.17}$$

defines the co-called *equilibrium constants* considered in the case of macroscopic equilibrium Gibbs states [12]. It follows from Eqs.(2.4) and (3.9) that the case $k_{xy} = 1$ for arbitrary $x, y \in X$ can take place only for the finite state space with the uniform Gibbs distribution, that is, if:

$$\forall x \in X, \ \pi(x) = Z^{-1} = \frac{1}{N}, \qquad N = \text{card}X. \tag{3.18}$$

The formula (3.15) has the form of the well-known law describing the frequency of the transition $x \to y$ in the theory of reaction dynamics and is applied, for example, to the description of the *thermally activated processes* [13, 14]. Basing on this observation, we can interpret $E_{xy}$ as the *energy barrier* between the states *x* and *y*, with own energies $E_x$ and $E_y$, respectively, whereas $U_{xy}$ can be interpreted as the *activation energy* of the change of states of the system leading from *x* to *y* [4]. Then the constant $v$ of Eq.(3.15) has the meaning of the effective frequency of efforts to overcome the energy barrier [13, 14].

Note that the *mean residence time* $\tau$ in the Gibbs state is given by

$$\tau = \int_X \tau(x)\pi(x)\,\mathrm{d}\mu(x), \tag{3.19}$$

where Eq.(2.37) was taken into account. If $\tau < \infty$, then we can assume that [4]

$$v = \frac{1}{\tau} \tag{3.20}$$

and, for the sufficiently large time $\tau$, say e.g. for

$$\tau \geq \tau_0 = \frac{1}{w_0} \tag{3.21}$$

in the case of Eq.(2.39), this nonequilibrium stationary state can be considered as a *metastable state*. For example, if we are dealing with *small nanostructured clusters*, it can be associated with the phenomenon of the existence of optimum size and shape leading to the most stable packing of their atoms [1].

### 4. Conditional entropy and external parameters

Let us consider a material *nonsingular semidynamical system* with the state space $X_\mu$ and such that every state $x \in X_\mu$ has its *own energy* $E_x \in \mathrm{R}_+$ independent of the dynamics of the system (Section 3). These energies can be dependent on a finite set *a* of parameters being external with respect to the system of particles under consideration [9]. For example if we are dealing with a system of material particles contained in a three-dimensional small convex figure *B* (constituting e.g. a *nanocluster* – [1]), then we can consider the triple $a = (\mathrm{V}, \mathrm{F}, \mathrm{M})$ of external parameters, where V is the volume of *B*, F is the surface field of the boundary $\partial B$ of *B* and M is the mean curvature of this boundary [1]. If we are dealing in a two-dimensional small convex figure *B* (e.g. a *graphene small cluster*), then the pair $a = (\mathrm{F}, \mathrm{M})$, where F is the area of *B*, $\mathrm{M} = (\pi/2)\mathrm{L}$ is the mean curvature of $\partial B$, and L is the perimeter of $\partial B$ [15], can be taken into account. Note that the volume parameter $a = \mathrm{V}$ is frequently discussed in classical thermodynamics of macroscopic systems [9, 16]. The corresponding *Gibbs distribution* $\pi$ (Section 3) depends then on these parameters, that is we have [9]

$$\pi(a,\theta,x) = Z(a,\theta)^{-1} \exp\left[-\frac{E_x(a)}{E_B}\right], \quad E_B = k_B\theta, \quad x \in X_\mu,$$

$$Z(a,\theta) = \int_X \exp\left[-\frac{e(a,x)}{E_B}\right] d\mu(x) < \infty, \quad e(a,x) = E_x(a),$$

(4.1)

where $\theta \in R_+$ is an absolute temperature and $k_B$ denotes the Boltzmann constant.

In statistical physics are considered the following equipotential sets of energy:

$$\Sigma_a(\varepsilon) = e_a^{-1}(\varepsilon), \quad e_a : X_\mu \to R_+,$$
$$\forall x \in X_\mu, \ e_a(x) = e(a,x).$$

(4.2)

It is admitted that the measure $\mu$ induces such measures $\mu_{a,\varepsilon}$ on the hypersurfaces $\Sigma_a(\varepsilon)$ in the state space that the generalized volume

$$\Omega(a,\varepsilon) = \text{vol}\Sigma_a(\varepsilon) > 0,$$

(4.3)

can be defined as a sufficiently smooth function of the parameters $a$ and $\varepsilon$. It follows from Eqs.(4.1)-(4.3) that the *conditional probability* density function $f(\varepsilon|a,\theta)$ of the energy distribution (with $a$ and $\theta$ keeping constant) [9]

$$f(\varepsilon|a,\theta) = \Omega(a,\varepsilon) Z(a,\theta)^{-1} \exp\left(-\frac{\varepsilon}{k_B\theta}\right),$$

(4.4)

describes then the distribution of values of own energies of states of a *system in a thermostat* [9, 16]. Introducing the *generalized free energy* of the system $F(a,\theta)$ by

$$F(a,\theta) = -E_B \ln Z(a,\theta),$$

(4.5)

and taking into account Eqs.(4.1) and (4.2), we can write the density function $f$ of Eq.(4.4) in the following form [9]:

$$f(\varepsilon|a,\theta) = \Omega(\varepsilon,a) \exp\left[\frac{F(a,\theta) - \varepsilon}{E_B}\right].$$

(4.6)

In statistical physics is considered also the *conditional free energy* $F(a,\theta|\varepsilon)$ of a system in the thermostat defined by the relation [9]:

$$f(\varepsilon|a,\theta) = \exp\left[\frac{F(a,\theta) - F(a,\theta|\varepsilon)}{E_B}\right],$$

(4.7)

or, according to Eq.(4.6), by the following formula:

`

$$F(a,\theta|\varepsilon) = \varepsilon - E_B \ln \Omega(a,\varepsilon), \qquad E_B = \theta k_B. \tag{4.8}$$

It follows from Eqs.(4.7) and (4.8) that [9]

$$f(\varepsilon_1|a,\theta) = f(\varepsilon_2|a,\theta)\exp\left[F(a,\theta|\varepsilon_2) - F(a,\theta|\varepsilon_1)\right]. \tag{4.9}$$

We see that a more probable state corresponds to a smaller value of the conditional free energy.

Introducing the *mean energy* $E(a,\theta)$ and the *entropy* $S(a,\theta)$ of the system as:

$$\begin{aligned} E(a,\theta) &= \int_X e(a,x)\pi(a,\theta,x)\,\mathrm{d}\mu(x), \\ S(a,\theta) &= \int_X s\big(\pi(a,\theta,x)\big)\,\mathrm{d}\mu(x), \end{aligned} \tag{4.10}$$

where it was denoted

$$s(z) = \begin{cases} -k_B z \ln z & \text{for } z > 0 \\ 0 & \text{for } z = 0, \end{cases} \tag{4.11}$$

we obtain the well-known thermodynamic relation:

$$F(a,\theta) = E(a,\theta) - \theta S(a,\theta). \tag{4.12}$$

Consequently, a (nonsingular) semidynamical system endowed with a distribution of own energies of its states can be considered as a thermodynamic system defined by Eqs.(4.1) and (4.10)-(4.12). The stationary states of such defined thermodynamic system are *Gibbs states* (Section 3).

Notice that Eqs.(4.8) and (4.10)-(4.12) suggest to define the *conditional entropy* $S(a,\varepsilon)$ of the system as [16] (cf. [9]):

$$S(a,\varepsilon) = k_B \ln \Omega(a,\varepsilon). \tag{4.13}$$

The *fundamental equation of thermodynamics* of reversible quasi-static transformations (of a system embedded in a thermostat) is the following [17]:

$$\mathrm{d}E = \theta\,\mathrm{d}S - A_k\,\mathrm{d}a^k, \qquad k = 1,2,\ldots n, \tag{4.14}$$

where $a = (a^k; k = 1,2,\ldots n)$ is the set of external parameters (called also *generalized coordinates*), $A_k, k = 1,2,\ldots n$, are the corresponding *generalized thermodynamic forces* and the relation

$$\mathrm{d}Q = \theta\,\mathrm{d}S, \tag{4.15}$$

where $Q$ is the so-called *heating* being a quantity describing the thermal influence of a thermostat on the system, is taken into account. Writing Eq.(4.14) in the following form:

$$dE = dQ - dA, \quad dA = A_k da^k, \tag{4.16}$$

and taking into account Eq.(4.10), we obtain that should be:

$$dE(a,\theta) = \int_X \left(\alpha_i(a,x)da^i\right)\pi(a,\theta,x)d\mu(x) + \int_X e(a,x)d_{a,\theta}\left(\pi(a,\theta,x)\right)d\mu(x), \tag{4.17}$$

where it was denoted

$$\alpha_i(a,x) = \frac{\partial e}{\partial a^i}(a,x), \quad d_{a,\theta}\pi = \frac{\partial \pi}{\partial a^i}da^i + \frac{\partial \pi}{\partial \theta}d\theta. \tag{4.18}$$

Defining the generalized thermodynamic forces $A_i$ as [9]:

$$A_i(a,\theta) = \int_X \alpha_i(a,x)\pi(a,\theta,x)d\mu(x), \tag{4.19}$$

and comparing Eq.(4.16) with Eqs.(4.17)-(4.19), we obtain the following representation of the change of heating:

$$dQ(a,\theta) = \int_X e(a,x)d_{a,\theta}\left(\pi(a,\theta,x)\right)d\mu(x). \tag{4.20}$$

Therefore, $dQ$ is defined by the change of Gibbs distribution due to the change of thermodynamic variables $a$ and $\theta$. Note that it follows from Eqs.(4.1), (4.5), (4.18) and (4.19) that [9]

$$A_i(a,\theta) = -\frac{\partial F}{\partial a^i}(a,\theta), \quad i = 1,2,...n. \tag{4.21}$$

Treating the entropy $S$ of Eq.(4.14) as a function of independent variables $E$ and $a^k$, we obtain that [17]

$$dS = \frac{1}{\theta}dE + \frac{1}{\theta}A_k da^k = \left(\frac{\partial S}{\partial E}\right)_a dE + \sum_{k=1}^{n}\left(\frac{\partial S}{\partial a^k}\right)_{E,a^{i\neq k}} da^k, \tag{4.22}$$

where the symbol $(\ )_z$ indicates that for the partial differentiation one should hold constant the variable $z$, and thus the following equations hold:

$$\frac{1}{\theta} = \left(\frac{\partial S}{\partial E}\right)_a, \quad A_k = \theta\left(\frac{\partial S}{\partial a^k}\right)_{E,a^{i\neq k}}. \tag{4.23}$$

We see that, from a thermodynamic point of view, the only requirement for the existence of a positive absolute temperature $\theta$ is that the entropy $S$ should be restricted to a monotonically increasing function of the internal energy $E$. Note that if we are dealing with the conditional entropy of Eq.(4.13), then

$$\left(\frac{\partial S}{\partial \varepsilon}\right)_a = k_B \Omega \left(\frac{\partial \Omega}{\partial \varepsilon}\right)_a, \tag{4.24}$$

where $\Omega(a,\varepsilon)d\varepsilon$, $a = \text{const.}$, can be interpreted as a number of admissible states of a system of $N$ particles with their own energies contained in the interval $[\varepsilon, \varepsilon + d\varepsilon]$ [16]. This number monotonically increases if $\varepsilon$ increases, i.e. we are dealing with states of the systems with no upper limit to the own energies of these states, e.g. for the kinetic energy of a gas molecule [16, 18] or in the case of harmonic oscillator [16]. Thus, for $\theta \in R_+$, the entropy of the system as well as its conditional entropy are monotonically increasing functions of the internal energy of the system. Nevertheless, some very peculiar systems that have energetic upper limits of their allowed states, are considered [16, 18]. The description of such systems in the framework of statistical physics based on the existence of Gibbs distribution, needs to introduce a negative absolute temperature [18]. In this case, according to Eqs.(4.26) and (4.27), the entropy of a thermodynamic system is not a monotonically increasing function of its internal energy. It ought to be stressed that the assumption relating to the sign of the absolute temperature is not explicitly made in thermodynamics. It is because such an assumption is not necessary in the derivation of many thermodynamic theorems [18].

It easy to see that if $\theta \in R_+$ and the *conditional probability* density function $f$ of Eq.(4.6) has an *extremum*, that is, there exists $\varepsilon = \varepsilon_m$ such that [9, 16]:

$$\left(\frac{\partial f}{\partial \varepsilon}\right)_{a,\theta}(\varepsilon_m|a,\theta) = 0, \tag{4.25}$$

or, equivalently, if

$$\left(\frac{\partial F}{\partial \varepsilon}\right)_a(a,\theta|\varepsilon_m) = 0, \tag{4.26}$$

then, according to Eq.(4.8), should be $\varepsilon_m = \varepsilon_m(a,\theta)$ where:

$$\left(\frac{\partial \Omega}{\partial \varepsilon}\right)_a (a,\varepsilon_m)_a = \frac{1}{k_B \theta} \Omega(a,\varepsilon_m), \tag{4.27}$$

and thus if the generalized volume $\Omega(a,\varepsilon)$, $a = \text{const.}$, increases if $\varepsilon$ increases:

$$\forall \varepsilon \in R_+, \left(\frac{\partial \Omega}{\partial \varepsilon}\right)_a (a,\varepsilon) > 0, \tag{4.28}$$

then $\theta > 0$. Consequently, according to Eq.(4.24), in this case the temperature $\theta$ is a positive scalar iff the conditional entropy $S$ of Eq.(4.13) is a monotonic function of the own energy $\varepsilon$, that is,

$$\forall \varepsilon \in R_+, \left(\frac{\partial S}{\partial \varepsilon}\right)_a (a,\varepsilon) > 0. \tag{4.29}$$

Notice that the density $f$ of the conditional probability has the *maximum* for the own energy $\varepsilon = \varepsilon_m(a,\theta)$ of Eq.(4.27) if

$$\left(\frac{\partial^2 f}{\partial \varepsilon^2}\right)_{a,\theta} (a,\theta|\varepsilon_m) < 0, \tag{4.30}$$

or, according to Eqs.(4.7) and (4.26), if:

$$\left(\frac{\partial^2 F}{\partial \varepsilon^2}\right)_{a,\theta} (a,\theta|\varepsilon_m) > 0. \tag{4.31}$$

## 5. Thermodynamically admissible Markov-type processes

Let us consider a *Markov-type material system* with its evolution governed by a nonsingular semidynamical system G such that the corresponding stochastic semigroup U[G] is generated by the Kolmogorow-type kinetic equation defined by (2.35)-(2.37), (3.15), and (alternatively) by (3.19)-(3.21) (Section 2). We will assume also that the system admits its treatment as a *small* thermodynamic system with a *diathermal* boundary (Sections 3 and 4). Let $X_\mu$ be the state space of this material system. A (stationary and nonequilibrium) *Gibbs state* (being perhaps a metastable state) of this system is described by Eqs.(3.1), (3.2), (3.6)-(3.11) (and, for example, fulfils the conditions (4.1), (4.10)-(4.12)). It ought to be stressed that the presented here approach can be also applied in the case of stochastic Markov processes (see Section 2 and [4]) and in the case of processes governed by the Chapman-Kolmogorov semigroup (Section 2).

We can now define the thermodynamically admissible *Markov-type process* of the evolution of the material system. First of all, such process should be consistent with the second law of thermodynamics. This condition can be formulated in the following way. Let us calculate functionals of the mean internal energy and the Boltzmann entropy (Section 3) along a trajectory of the *Markov-type semigroup* $U[G] = \{U_t : D_\mu(X) \to D_\mu(X), \; t \in T\}$ corresponding to G (Section 2), that is, defined by Eqs.(2.1), (2.2), (2.9)-(2.12), (2.20), (2.29) (in the case of deterministic systems; say e.g. defined by Eq.(2.3)) or by Eqs.(2.16)-(2.20) (in the case of stochastic systems):

$$\forall t \in T, \; E(t) = E(U_t p), \; S(t) = S(U_t p), \tag{5.1}$$

where $p \in D_\mu(X)$ is an initial stationary distribution. Since the environment of the considered system is a thermostat (Section 3), we can consider *nonequilibrium states* of this system with its constant temperature $\theta > 0$ defined by this thermostat. This makes possible the following extension of the definition (3.11) of the generalized free energy (corresponding to the Gibbs state) to the case of *nonequilibrium isothermal processes*:

$$\forall t \in T, \, F(t) = F(U_t p),$$
$$\forall p \in D_\mu(X), \, F(p) = E(p) - \theta S(p), \quad \theta = \frac{E_B}{k_B}, \quad (5.2)$$

where $E_B$ is a *characteristic energy* of the system associated with the considered thermal phenomena. Note that the similar definition of the nonequilibrium free energy is formulated in order to describe translational Brownian motion in an equilibrium medium treated as a thermostat [9].

It follows from Eqs.(5.1) and (5.2) that

$$dS = \theta^{-1}(dE - dF), \quad (5.3)$$

where it was denoted $dh = \dot{h} dt$ and $\dot{h} = dh/dt$ for a differentiable function $h: T \to \mathbb{R}$. The Boltzmann entropy increment $\delta_e S$ due to the interaction of the system with its environment is given by:

$$\delta_e S = \theta^{-1} \delta Q, \quad (5.4)$$

where $\delta Q$ is the heat increment. So, the Boltzmann entropy increment $\delta_i S$ due to the existence of thermodynamically irreversible processes in the system can be calculated from Eqs (5.3) and (5.4):

$$\delta_i S = dS - \delta_e S = -\theta^{-1}(dF + dE - \delta Q). \quad (5.5)$$

The considered Markov-type process will be consistent with the *second law thermodynamics* if and only if

$$\delta_i S \geq 0. \quad (5.6)$$

The interaction of the system with its environment has only the thermal character (Section 3) if and only if the *first law of thermodynamics* takes the following form:

$$dE = \delta Q. \quad (5.7)$$

It follows from Eqs.(5.2) and (5.5)-(5.7) that the considered Markov-type processes can be treated, in the diathermal and isothermal conditions, as *thermodynamically admissible* if and only if the generalized free energy functional (Section 3) is non-increasing along the trajectories of the corresponding Markov-type semigroups, that is,

$$\forall t \in T, \ \dot{F}(t) \leq 0. \tag{5.8}$$

These trajectories are defined by Eq.(2.19), by the *Kolmogorov-type kinetic equation* (2.36) with the transition intensities given by Eq.(3.15) (and, perhaps, additionally by Eqs.(3.19) and (3.20)), and by the rule:

$$\forall x \in X, \ p_t(x) = \overline{p}(x,t), \ t \in T = \mathrm{R}_+. \tag{5.9}$$

If the stochastic semigroup is the *Chapman-Kolmogorov semigroup* (Section 2), then Eq.(2.25) should be additionally taken into account. It is, for example, the case of Markov chains [4, 11].

It seems physically reasonable to distinguish the class of thermodynamically admissible *irreversible Markov-type processes* consistent with the existence of Gibbs states. It can be formulated as a *relaxation postulate* stating that the Markov-type irreversible processes relax, independently of the choice of the initial condition, to the univocally defined Gibbs state (being perhaps a metastable state). Note that since the Gibbs distribution $\pi$ of Eqs.(3.9) or (4.1) fulfils identically Eqs.(2.36) with $W(x,y)$ given by Eq.(3.15) and $p = \pi$, the condition (2.22) is fulfilled. Moreover, according to this postulate, the condition (2.23) of asymptotical stability should be satisfied. The relaxation postulate is fulfilled, for example, in the case of thermodynamically admissible Markov chains with the continuous time and the finite state space [4, 11] (see Section 2).

The relaxation postulate should be treated as the additional thermodynamic postulate that defines the notion of Gibbs states more precisely (cf. [4] and [19]). However, it ought to be stressed that in many cases it is difficult to prove that the probabilistic representation of a process in the state space fulfills this postulate. For example, in the case of the Chapman-Kolmogorov semigroup (Section 2), the condition (2.22) means that should be:

$$\int_X K_t(x,y)\pi(y)\mathrm{d}\mu(y) = \pi(x), \tag{5.10}$$

for every $x \in X$ and $t \in T$. Note also that the Gibbs distribution fulfils the Liouville equation (2.14) but the Hamiltonian system does not relax to the Gibbs state.

## 6. Conclusions and remarks

The notion of *irreversibility* is based on the endowing of the time axis with the distinguished "forward" orientation (see Section 1). In the classical approach, based on the Hamiltonian microstate dynamics (see remarks in Sections 2 and 3), the time is considered as oriented only when the collective ("macroscopic" or "mesoscopic") properties of material systems are described, and remains non-oriented when the material system is analyzed on the micro-level (it is because the conservative Hamiltonian dynamics does not distinguish any direction of time). This duality in treating the time can be eliminated, for example, in the case of material systems (deterministic as well as stochastic) revealing Markov-type irreversible evolution (Section 2). These evolution processes are defined as nonsingular semidinamical systems possessing a probabilistic representation described by the Kolmogorov-type kinetic equation (2.36) introduced in Section 2. The description of the irreversible evolution of the system can be reduced then to the formulation of the physical hypothesis about the form of the transition probabilities (Section 2). In Section 3 are considered stationary states of not closed Markov-type small material systems (characterized, for example, by suitably selected external parameters – see remarks at the beginning of Section 4) and it is shown that the Gibbs distribution (Sections 3 and 4) fulfils the Kolmogorov-type kinetic equation if an analogue of the so called condition of microscopic reversibility (called also the condition of detailed balance) is satisfied. It turns out that the proposed theory describes then thermally activated collective processes (Section 3). It is shown also that the Markov-type processes can be treated, in the diathermal and isothermal conditions, as thermodynamically admissible if and only if the generalized free energy functional (Section 3) is non-increasing along the trajectories of the corresponding Markov-type semigroups (Section 5).

The existence of the above-mentioned diathermal conditions enables to consider a small material system endowed, due to its contact with a thermostat, with the positive absolute temperature (Sections 3 and 4) and, consequently, enables to introduce the notion of generalized free energy of nonequilibrium isothermal processes (Sections 3 and 5). Thus, since the considered dynamics of small material systems takes place in isothermal conditions, we are dealing rather with the description of "*isothermal collective properties*" of these systems consistent with thermodynamical rules than with the standard statistical thermodynamics of macroscopic material systems (see also the case of "*nanothermomechanics*" [1] – a nanoscale isothermal counterpart of analytical mechanics of affinely-rigid macroscopic bodies [21] consistent with the phenomenological thermodynamics). For example, in Section 5, in addition to the first and second laws of thermodynamics treated as thermodynamic constraints of Markov-type material systems, an additional isothermal constraint of these systems (called the *relaxation postulate*) is formulated and discussed.

Note that the presented approach admits the consistency with the so-called *Prigogine's selection rule*, according to which only these probabilistic representations of dynamics that are directed "forward" describe physically realizable states [20]. Prigogine assumes additionally that this selection rule cannot be derived from dynamics in this sense that it is not related with the existence of any new interactions not yet taken into account. In this approach, the time asymmetry exhibits itself on the microlevel in the form of internal random nature of the system (that is, independently of any hidden variables). According to the Prigogine's point of view, the time asymmetry should be universal, that is, it should take place in all dynamical theories: in the classical mechanics as well as in the quantum mechanics (and in relativistic theories. This point of view can be useful in the case of the description of effective thermomechanical properties of bulk nanostructured clusters (see, for example, nanothermomechanics of bulk nanoclusters considered in [1]) as well as in the case of low-dimensional material systems (see, for example, [1] – remarks concerning planar graphene sheets, [22], [23], and [24] where the "*isothermal geometry*" of corrugated graphene sheets is formulated). Namely, these material systems are sufficiently small (in the case of bulk nanoclusters) or are sufficiently thin (in the case of graphene sheets) so they are not completely free of quantum effects and thus, they not simply obey the classical physics governing the macroworld (Section 1). Moreover, in the case of graphene sheets, it ought to be taken into account that graphene is intrinsically not flat and corrugated randomly ([25] and references therein). Consequently, at least in the case of low-dimensional small material systems (deterministic or stochastic – see Section 2 and assumptions at the beginning of Section 5), the second law of thermodynamics (Section 5) attains the status of the fundamental law of dynamics of these systems. If so, the problem of the existence of a mapping realizing the probabilistic representation of the microstate dynamics of the above mentioned small material systems becomes of fundamental importance.

**Acknowledgement**

This paper contains results obtained within the framework of the research project N N501 049540 financed from Scientific Research Support Fund in 2011-2014 The author is greatly indebted to the Polish Ministry of Science and Higher Education for this financial support.